\RequirePackage{lineno} 
\documentclass[aps,prc,superscriptaddress,nofootinbib,showpacs,floatfix,singlecolumn]{revtex4}
\usepackage{graphicx} 
\usepackage{float} 
\usepackage{amssymb}
\usepackage{url}
\usepackage[colorlinks,linkcolor=blue,citecolor=blue,filecolor=black,urlcolor=blue]{hyperref}
\usepackage{cancel}
\usepackage{MnSymbol}
\usepackage{multirow}
\usepackage{bm}
\newcommand{\scn}{\mathrm{SC}(n,m)}
\newcommand{\sca}{\mathrm{SC}(2,4)}
\newcommand{\scb}{\mathrm{SC}(2,3)}

\newcommand{\lr}[1]{\left\langle #1\right\rangle}
\newcommand{\llrr}[1]{\left\llangle #1\right\rrangle}
\newcommand{\pT} {\ensuremath{p_{\mathrm{T}}}}

\newcommand{\pp}{\mbox{$pp$}}
\newcommand{\pPb}{\mbox{$p$+Pb}}

\newcommand{\pA}{\mbox{$p$+A}}

\newcommand{\nch}{\mbox{$N_{\mathrm{ch}}$}}
\newcommand{\nchb}{\mbox{$N_{\mathrm{ch}}^{\textrm{sel}}$}}

\newcommand{\sqrtsnn}{\mbox{$\sqrt{s_{\mathrm{NN}}}$}}
\newcommand{\Dphi}{\mbox{$\Delta \phi$}}
\newcommand{\Deta}{\mbox{$\Delta \eta$}}

\usepackage{color}
\usepackage{harpoon}
\definecolor{my}{rgb}{1, 0, 0}
\begin{document} 
\title{Importance of non-flow in mixed-harmonic multi-particle correlations in small collision systems}
 \newcommand{\sunysb}{Department of Chemistry, Stony Brook University, Stony Brook, NY 11794, USA}
 \newcommand{\bnl}{Physics Department, Brookhaven National Laboratory, Upton, NY 11796, USA}{
 \newcommand{\nbi}{Niels Bohr Institute, University of Copenhagen, Blegdamsvej 17, 2100 Copenhagen, Denmark}

\author{Peng Huo}\affiliation{\sunysb}
\author{Katar\'{i}na~Gajdo\v{s}ov\'{a}}\affiliation{\nbi}
\author{Jiangyong Jia}\email[Correspond to\ ]{jjia@bnl.gov} \affiliation{\sunysb}\affiliation{\bnl}
\author{You Zhou}\email[Correspond to\ ]{you.zhou@cern.ch}\affiliation{\nbi}
\date{\today}

\begin{abstract}
Recently CMS Collaboration measured  mixed-harmonic four-particle azimuthal correlations, known as symmetric cumulants $\scn$, in $pp$ and $\pPb$ collisions, and interpreted the non-zero $\scn$ as evidence for long-range collectivity in these small collision systems.  Using the PYTHIA and HIJING models which do not have genuine long-range collectivity, we show that the CMS results, obtained with standard cumulant method, could be dominated by non-flow effects associated with jet and dijets, especially in $pp$ collisions. We show that the non-flow effects are largely suppressed using the recently proposed subevent cumulant methods by requiring azimuthal correlation between two or more pseudorapidity ranges. We argue that the reanalysis of $\scn$ using the subevent method in experiments is necessary before they can used to provide further evidences for a long-range multi-particle collectivity and constraints on theoretical models in small collision systems.
\end{abstract}
\maketitle
\section{Introduction}\label{sec:1}

Measurements of two-particle angular correlation in small collision systems, such as $pp$ or $\pA$, have revealed the ridge phenomena~\cite{CMS:2012qk,Abelev:2012ola,Aad:2012gla,Aad:2014lta,Khachatryan:2015waa}: enhanced production of pairs at small azimuthal angle separation, $\Dphi$, extended over wide range of pseudorapidity separation $\Deta$. The azimuthal structure of the ridge is often characterized by a Fourier series $dN_{\mathrm {pair}}/d\Dphi \sim 1+2 \Sigma v_n^2\cos (n\Dphi)$, and studied as a function of charged particle multiplicity $\nch$. The $v_n$ denotes the anisotropy coefficients for single particle distribution, with $v_2$ being the largest followed by $v_3$. The ridge reflects multi-parton dynamics at early time of the collision and has generated significant interests in high-energy physics community. One key question concerning the ridge is the timescale for the emergence of the long-range multi-particle collectivity, whether it reflects initial momentum correlation from gluon saturation effects~\cite{Dusling:2013qoz} or it reflects a final-state hydrodynamic response to the initial transverse collision geometry~\cite{Bozek:2013uha}. 

More insights about the ridge is obtained via multi-particle correlation technique, known as cumulants, involving four or more particles~\cite{Borghini:2000sa,Bilandzic:2010jr,Bilandzic:2013kga,jjia}. The multi-particle cumulants probe the event-by-event fluctuation of $v_n$, $p(v_n)$, as well as the correlation between $v_n$ of  different order, $p(v_n, v_m)$. For example, four-particle cumulant $c_n\{4\}=\lr{v_n^4}-2\lr{v_n^2}^2$ constrains the width of $p(v_n)$~\cite{Borghini:2000sa}, while four-particle symmetric cumulants $\scn=\lr{v_n^2v_m^2}-\lr{v_n^2}\lr{v_m^2}$ quantifies the lowest-order correlation between $v_n$ and $v_m$~\cite{Bilandzic:2013kga}. 

The main challenge in the study of azimuthal correlations in small systems is how to distinguish long-range ridge correlations from ``non-flow'' correlations such as resonance decays, jets, or dijet production. In A+A collisions, non-flow is naturally suppressed due to large particle multiplicity, i. e. non-flow contribution scales as $1/\nch$ and  $1/N_{\mathrm {ch}}^3$ for the two- and four-particle cumulants, respectively~\cite{Borghini:2001vi}. In small systems, however, non-flow can be large due to their much smaller $\nch$ values, and one has to empoly new methods that explicitly exploit the long-range nature of the collectivity in $\eta$: For two-particle correlations, the non-flow is suppressed by requiring a large $\Deta$ gap and a peripheral subtraction procedure~\cite{Abelev:2012ola,Aad:2012gla,Aad:2014lta,Aad:2015gqa,Aaboud:2016yar,Khachatryan:2016txc}. For multi-particle cumulants, the non-flow can be suppressed by requiring correlation between particles from different subevents separated in $\eta$, while keeping the genuine long-range multi-particle correlations associated with the ridge. This so-called subevent method~\cite{jjia} has been shown to be necessary to obtain a reliable $c_n\{4\}$~\cite{Aaboud:2017blb}, while the $c_2\{4\}$ based on the standard cumulant method~\cite{Khachatryan:2016txc,Aaboud:2017acw} are contaminated by non-flow correlations over the full $\nch$ range in $\pp$ collisions and the low $\nch$ region in $\pA$ collisions.

Recently CMS Collaboration also released measurements of $\scb$ and $\sca$ in $pp$ and $\pPb$ collisions, based on the standard cumulant method~\cite{Sirunyan:2017uyl}. However, since these observables have much smaller signal than $c_2\{4\}$, they are expected to be even more susceptible to non-flow effects. Therefore, more precise study of the influence of non-flow effects to these observables is required before any interpretation of the experimental measurements. Event generators such as PYTHIA8~\cite{Sjostrand:2007gs} and HIJING~\cite{Gyulassy:1994ew}, which contain only non-flow correlations, are perfect test-ground for estimating the influence of non-flow to symmetric cumulants in small systems, which is the focus of this paper. Using a PYTHIA8 simulation of $pp$ collisions and HIJING simulation of $\pPb$ collisions, we demonstrate that $\scn$ based on the standard method is dominated by non-flow in $pp$ collisions, and is contaminated by non-flow in $\pPb$ collisions. We show that reliable $\scn$ measurements can be obtained using three-subevent or four-subevent methods, which therefore should be the preferred methods for analyzing multi-particle correlations in small systems. 

\section{Symmetric cumulants}\label{sec:2}
The framework for the standard cumulant is described in Refs.~\cite{Bilandzic:2010jr,Bilandzic:2013kga}, which was recently extended to the case of subevent cumulants in Ref.~\cite{jjia,Gajdosova:2017fsc}. The four-particle symmetric cumulants $\scn$ are related to two- and four-particle azimuthal correlations for flow harmonics of order $n$ and $m$, $n\neq m$ as:
\begin{eqnarray}\label{eq:1}
&&\lr{\{4\}_{n,m}}= \lr{{\mathrm{e}}^{{\rm i}n(\phi_1-\phi_2)+{\rm i}m(\phi_3-\phi_4)}}\;,\;\lr{\{2\}_{n}}=\lr{{\mathrm{e}}^{{\rm i}n(\phi_1-\phi_2)}}\;,\; \lr{\{2\}_{m}}=\lr{{\mathrm{e}}^{{\rm i}m(\phi_1-\phi_2)}}\;,\\
&&\scn = \llrr{\{4\}_{n,m}}-\llrr{\{2\}_{n}}\llrr{\{2\}_{m}}=\llrr{{\mathrm{e}}^{{\rm i}n(\phi_1-\phi_2)+{\rm i}m(\phi_3-\phi_4)}}- \llrr{{\mathrm{e}}^{{\rm i}n(\phi_1-\phi_2)}} \llrr{{\mathrm{e}}^{{\rm i}m(\phi_1-\phi_2)}}\;.
\end{eqnarray}
One firstly averages all distinct quadruplets or pairs in one event to obtain $\lr{\{4\}_{n,m}}$, $\lr{\{2\}_{n}}$ and $\lr{\{2\}_{m}}$, then average over an event ensemble to obtain $\llrr{\{4\}_{n,m}}$, $\llrr{\{2\}_{n}}$, $\llrr{\{2\}_{m}}$ and $\scn$. In the absence of non-flow correlations, $\scn$ measures the correlation between event-by-event fluctuations of $v_n$ and $v_m$:
\begin{eqnarray}\label{eq:2}
\scn_{\mathrm{flow}}=\lr{v_n^2v_m^2}-\lr{v_n^2}\lr{v_m^2}
\end{eqnarray}

In the standard cumulant method, all quadruplets and pairs are selected using the entire detector acceptance. To suppress the non-flow correlations that typically involve particles emitted within a localized region in $\eta$, the particles can be grouped into several subevents, each covering a non-overlapping $\eta$ interval. The multi-particle correlations are then constructed by correlating particles between different subevents, further reducing non-flow correlations. 

Specifically, in the two-subevent cumulant method, the entire event is divided into two subevents, labeled as $a$ and $b$, for example according to $-\eta_{\rm{max}}<\eta_a<0$ and $0<\eta_b<\eta_{\rm{max}}$. The symmetric cumulant is defined by considering all quadruplets comprised of two particles from each subevent, or pairs comprised of one particle from each subevent:
\begin{eqnarray}\label{eq:3}
\scn_{\mathrm{2-sub}} = \llrr{{\mathrm{e}}^{{\rm i}n(\phi_1^a-\phi_2^b)+{\rm i}m(\phi_3^a-\phi_4^b)}}- \llrr{{\mathrm{e}}^{{\rm i}n(\phi_1^a-\phi_2^b)}} \llrr{{\mathrm{e}}^{{\rm i}m(\phi_1^a-\phi_2^b)}}\;,
\end{eqnarray}
where the superscript or subscript $a$ ($b$) indicates particles chosen from the subevent $a$ ($b$). The two-subevent method suppresses correlations within a single jet (intra-jet correlations), since each jet usually emits particles to one subevent.

Similarly for the three-subevent and four-subevent methods, the $|\eta|<\eta_{\rm{max}}$ range is divided into three or four equal ranges, and are lablled as $a$, $b$ and $c$ or $a$, $b$, $c$ and $d$, respectively. The corresponding symmetric cumulants are defined as:
\begin{eqnarray}\label{eq:4}
\scn_{\mathrm{3-sub}} = \llrr{{\mathrm{e}}^{{\rm i}n(\phi_1^a-\phi_2^b)+{\rm i}m(\phi_3^a-\phi_4^c)}}- \llrr{{\mathrm{e}}^{{\rm i}n(\phi_1^a-\phi_2^b)}} \llrr{{\mathrm{e}}^{{\rm i}m(\phi_1^a-\phi_2^c)}}\\
\scn_{\mathrm{4-sub}} = \llrr{{\mathrm{e}}^{{\rm i}n(\phi_1^a-\phi_2^b)+{\rm i}m(\phi_3^c-\phi_4^d)}}- \llrr{{\mathrm{e}}^{{\rm i}n(\phi_1^a-\phi_2^b)}} \llrr{{\mathrm{e}}^{{\rm i}m(\phi_1^c-\phi_2^d)}}
\end{eqnarray}
Since the two jets in a dijet event usually produce particles in at most two subevents, the three-subevent and four-subevent method further suppresses inter-jet correlations associated with dijets. Furthermore, four-subevent suppresses possible three-jet correlations, although such contributions are expected to be small. To enhance the statistical precision, the $\eta$ range for subevent $a$ is also interchanged with that for subevent $b$, $c$ or $d$, which results in three independent $\scn_{\mathrm{3-sub}}$ and three independent $\scn_{\mathrm{4-sub}}$. They are averaged to obtain the final result.

\section{Model setup}\label{sec:3}
To evaluate the influence of non-flow to $\scn$ in the standard and subevent method, the PYTHIA8 and HIJING models are used to generate $\pp$ events at $\sqrt{s} = 13$ GeV and $\pPb$ events at $\sqrtsnn = 5.02$ TeV, respectively. These models contain significant non-flow correlations from jets, dijets, and resonance decays, which are reasonably tuned to describe the data, such as $\pT$ spectra, $\nch$ distributions. Multi-particle cumulants based on the standard method as well as subevent methods are calculated as a function of charged particle multiplicity $\nch$. To make the results directly comparable to the CMS measurement~\cite{Sirunyan:2017uyl}, the cumulant analysis is carried out using charged particles in $|\eta|<\eta_{\rm{max}}=2.5$ and several $\pT$ ranges, and the $\nch$ is defined as the number of charged particles in $|\eta|<2.5$ and $\pT>0.4$ GeV.

The symmetric cumulants are calculated in several steps using charged particles with $|\eta|<2.5$, similar to Refs.~\cite{jjia,Aaboud:2017blb}. Firstly, the multi-particle correlators $\lr{\{2k\}}$ with $k=1,2$ (indexes $n$ and $m$ are dropped for simplicity) in Eq.~\ref{eq:1} are calculated for each event from particles in one of the two $\pT$ ranges, $0.3<\pT<3$ GeV and $0.5<\pT<5$ GeV, and the number of charged particle in this $\pT$ range, $\nchb$, is calculated. Note that $\nchb$ is not the same as $\nch$ defined earlier due to different $\pT$ ranges used. Secondly, $\lr{\{2k\}}$ are averaged over events with the same $\nchb$ to obtain $\llrr{\{2k\}}$ and $\scn$. The $\scn$ values calculated for unit $\nchb$ bin are then combined over broader $\nchb$ ranges of the event ensemble to obtain statistically significant results. Finally, the $\scn$ obtained for a given $\nchb$ are mapped to given $\lr{\nch}$ to make the results directly comparable to the CMS measurements~\cite{Sirunyan:2017uyl}.

To further study the influence of non-flow fluctuations associated with multiplicity fluctuations, several other $\pT$ ranges, different from those used for $\lr{\{2k\}}$, are also used to calculated $\nchb$. The results from this study are discussed in Appendix~\ref{sec:6}.

\section{Results}\label{sec:4}
First we calculate the $\sca$ and $\scb$ from PYTHIA and HIJING using the standard cumulant method and compare them with the CMS $\pp$ and $\pPb$ data for charged particles. The same $\pT$ selection, $0.3<\pT<3$ GeV, is used to calculate the cumulants as well as to select the event class $\nchb$.

The comparison is shown in Figure~\ref{fig:1}. The results from models are non-zero and they decrease as a function of $\nch$ similar to the data, indicating that the data may have significant non-flow contributions. In $pp$ collisions as shown in the left panel, both $\sca$ and $\scb$ from the PYTHIA8 model are larger than the data, suggesting that either PYTHIA8 overestimates the non-flow contribution in $\scn$ or the flow correlation signals are negative. In $\pPb$ collisions as shown in the right panel, $\scn$ from the HIJING model are larger than (for $\scb$) or roughly comparable (for $\sca$) with the data for $\nch<70$, but their magnitudes are much smaller than the data for $\nch>100$. This implies that the influence of the non-flow is subdominant in $\pPb$ collisions, about 20\% or less, at large $\nch$ region, but it still dominates the small $\nch$ region. 

\begin{figure}[h!]
\begin{center}
\includegraphics[width=1\linewidth]{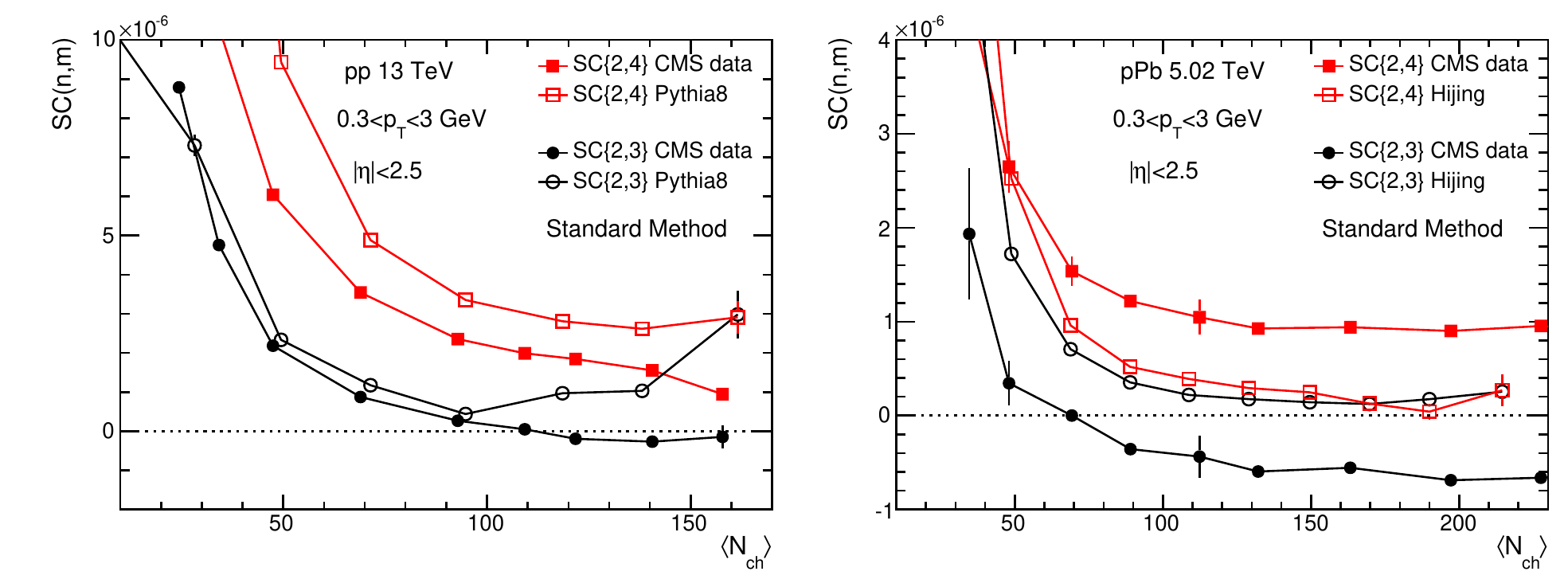}
\end{center}
\caption{\label{fig:1} The $\scn$ calculated for charged particles with $0.3<\pT<3$~GeV with the standard cumulant method in 13~TeV $\pp$ collisions (left panel) and 5.02 TeV $\pPb$ collisions (right panel) compared between data (solid symbols) and Monte Carlo models (open symbols).}
\end{figure}

\begin{figure}[h!]
\begin{center}
\includegraphics[width=1\linewidth]{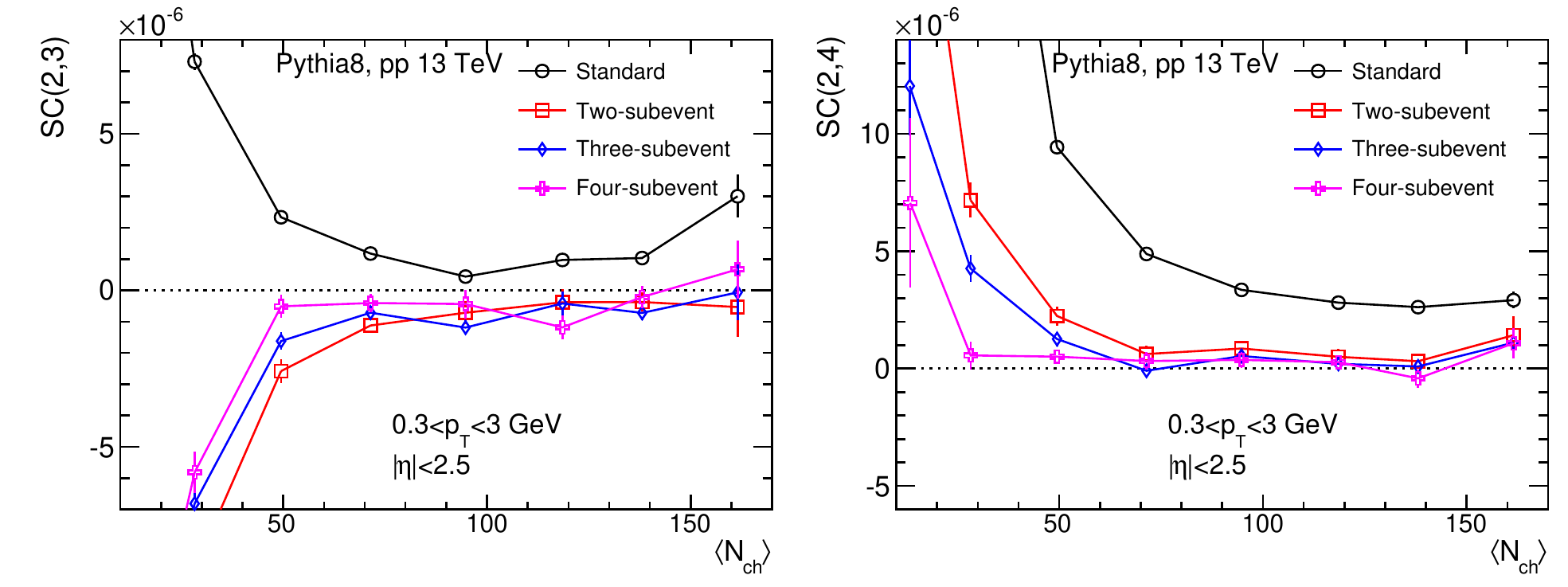}
\end{center}
\caption{\label{fig:2} The $\scb$ (left panel) and $\sca$ (right panel)  in $0.3<\pT<3$ GeV and $|\eta|<2.5$ as a function of $\nch$ obtained from 13 TeV $pp$ PYTHIA 8 simulations using the standard cumulant, two-subevent, three-subevent and four-subevent methods.}
\end{figure}
The comparison shown in Figure~\ref{fig:1} suggests that the symmetric cumulants measured with the standard method are strongly biased by non-flow correlations in $pp$ collisions over the full $\nch$ range and in $\pPb$ collisions at low $\nch$ region.  On the other hand, the non-flow correlations are expected to be greatly suppressed in the subevent methods. Figures~\ref{fig:2} and \ref{fig:3} show $\scn$ obtained from various methods in $\pp$ collisions for charged particles in $0.3<\pT<3$ GeV and $0.5<\pT<5$ GeV, respectively. The same $\pT$ selections are used to calculate the cumulants as well as to select the event class.
\begin{figure}[h!]
\begin{center}
\includegraphics[width=1\linewidth]{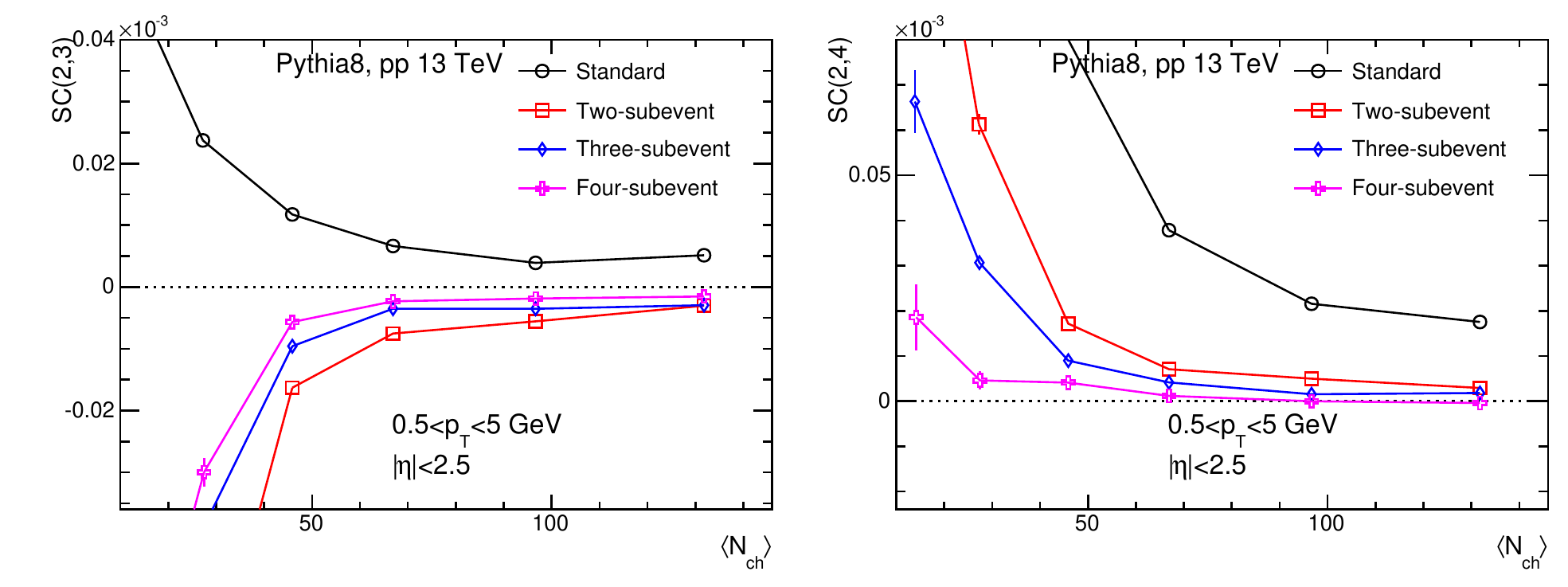}
\end{center}
\caption{\label{fig:3} The $\scb$ (left panel) and $\sca$ (right panel) for charged particles in $0.5<\pT<5$ GeV and $|\eta|<2.5$ as a function of $\nch$ obtained from 13 TeV $pp$ PYTHIA 8 simulations using the standard cumulant, two-subevent, three-subevent and four-subevent methods.}
\end{figure}

Figures~\ref{fig:2} and \ref{fig:3} show that the values of $\scn$ from subevent methods are much smaller than those from the standard method. In particular, the four-subevent method gives results that are closest to 0, followed by the three-subevent method and then the two-subevent method. Since non-flow contributions are known to increase with $\pT$, such hierarchy between different methods are more clearly revealed in Figure~\ref{fig:3} than in Figure~\ref{fig:2}. It is also interesting to note that the values of $\scb$ is negative in the subevent methods, and can't be fully suppressed to zero even in the four-subevent method. The sign-change of $\scb$ between the standard and two-subevent can be understood as the interplay between the inter-jet and intra-jet correlations: while the inter-jet correlation gives a positive contribution to $\scb$, the intra-jet correlation from dijets gives a negative contribution. The $\scb$ in standard method is positive because the inter-jet correlation dominates over the intra-jet contribution. However since the dijet contributions are further suppressed in the three-subevent and four-subevent methods, the residual negative $\scb$ in these two methods suggest the existence, in PYTHIA8 and HIJING, of a small long-range non-flow source that correlate between the $2^{\mathrm{nd}}$ and $3^{\mathrm{rd}}$ harmonics.

Similar observations are found in $\pPb$ collisions as shown in Figure~\ref{fig:4}, i.e. results from the subevent methods are closer to zero than those from the standard method. However, due to a much smaller non-flow in $\pPb$ collisions ($\sim$ ten times smaller than $pp$ at comparable $\nch$ for $\nch>100$),  the precision of the simulation does not allow a clear separation between different subevent methods. This also implies that we can already obtain reliable $\scn$ as soon as the subevent method is applied.
\begin{figure}[h!]
\begin{center}
\includegraphics[width=1\linewidth]{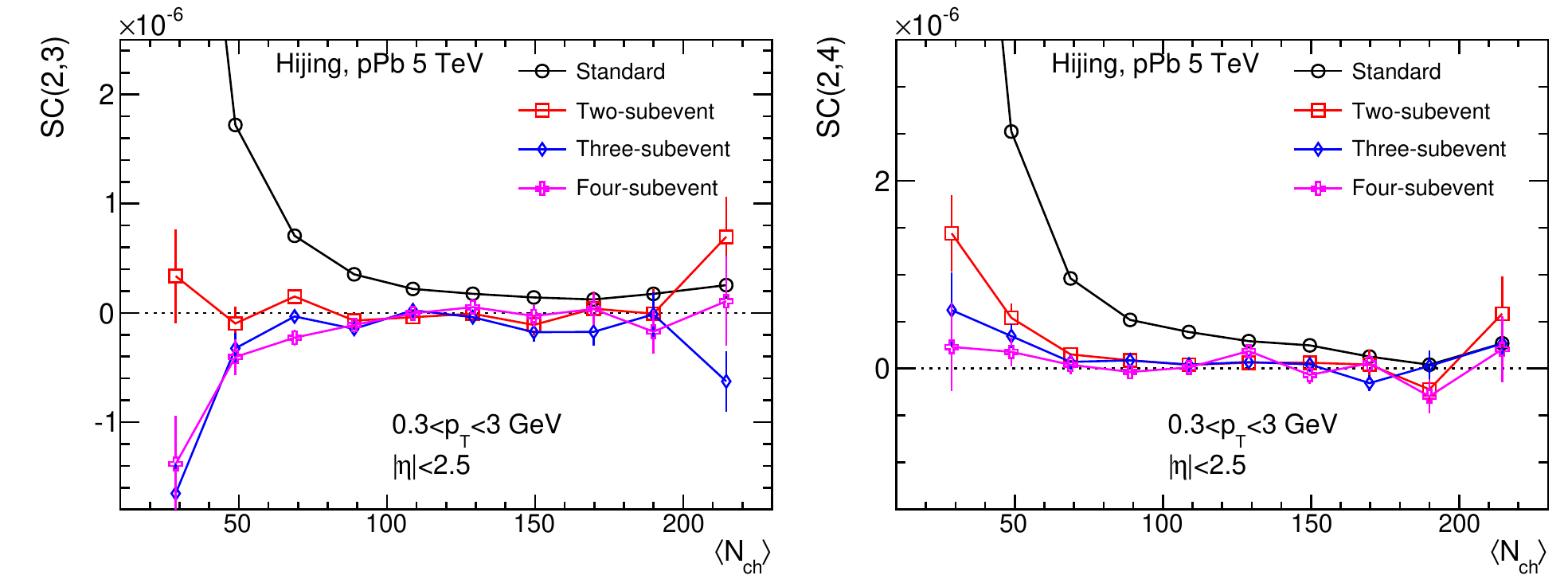}
\end{center}
\caption{\label{fig:4} The $\scb$ (left panel) and $\sca$ (right panel)  in $0.3<\pT<3$ GeV and $|\eta|<2.5$ as a function of $\nch$ obtained from 5.02 TeV $\pPb$ HIJING simulations using the standard cumulant, two-subevent, three-subevent and four-subevent methods.}
\end{figure}

\section{Summary}\label{sec:5}
Multi-particle azimuthal correlation between different flow harmonics $v_n$ and $v_m$, known as symmetric cumulants $\scn$, has been used to study the nature of the long-range ridge in $pp$ and $\pPb$ collision. Using the PYTHIA and HIJING models which contains only non-flow correlations, we show that recently measured $\sca$ and $\scb$, by the CMS Collaboration via the standard cumulant method, are likely contaminated by non-flow associated with jet and dijets. By requiring azimuthal correlation between multiple pseudorapidity $\eta$ ranges, we show that calculations using the recently proposed subevent cumulant methods are much less sensitive to these non-flow sources. Although the subevent methods can suppressed $\sca$ to nearly zero in high-multiplicity $\pp$ and $\pPb$ collisions, the $\scb$ from subevent methods still shows a small but negative correlation in these collisions. These studies suggest that the measurements of $\scn$ need to be redone with the subevent methods, before any physics conclusion related to long-range collectivity can be drawn.

J.J and P.H's research is supported by National Science Foundation under grant number PHY-1613294. Y.Z and K.S's research is supported by the Danish Council for Independent Research, Natural Sciences, the Danish National Research Foundation (Danmarks Grundforskningsfond) and the Carlsberg Foundation (Carlsbergfondet).
\section*{Appendix: sensitivity to event class definition}\label{sec:6}
Another way to quantify the influence of the non-flow in the cumulant method is to study the sensitivity of $\scn$ on the choice of $\nchb$. Previous studies shows that different $\nchb$ leads to drastically change the nature of the non-flow fluctuations, leading to different cumulant results. Following the example of Ref.\cite{jjia,Aaboud:2017blb}, the impact of non-flow fluctuations to $\scn$ are probed by varying the $\pT$ requirements used to define $\nchb$ as follows: When $\lr{\{2k\}}$ is calculated in the range $0.3<\pT<3$ GeV, $\nchb$ is evaluated in four different track $\pT$ ranges: $0.3<\pT<3$ GeV, $\pT>0.2$ GeV, $\pT>0.4$ GeV and $\pT>0.6$ GeV.  When $\lr{\{2k\}}$ is calculated in $0.5<\pT<5$ GeV, $\nchb$ is evaluated in four different track $\pT$ ranges: $0.5<\pT<5$ GeV, $\pT>0.2$ GeV, $\pT>0.4$ GeV and $\pT>0.6$ GeV. The  $\scn$ values obtained for a given $\nchb$ are mapped to given $\nch$, so that $\scn$ obtained for different $\nchb$ can be compared using a common $x$-axis defined by $\nch$.

The results of this study are shown in Figure~\ref{fig:1a} and \ref{fig:2a} for $pp$ and $\pPb$ collisions, respectively. A strong sensitivity of $\scn$ on $\nchb$ is observed in the standard method. But such sensitivity is greatly reduced in the subevent method.

\begin{figure}[h!]
\begin{center}
\includegraphics[width=1\linewidth]{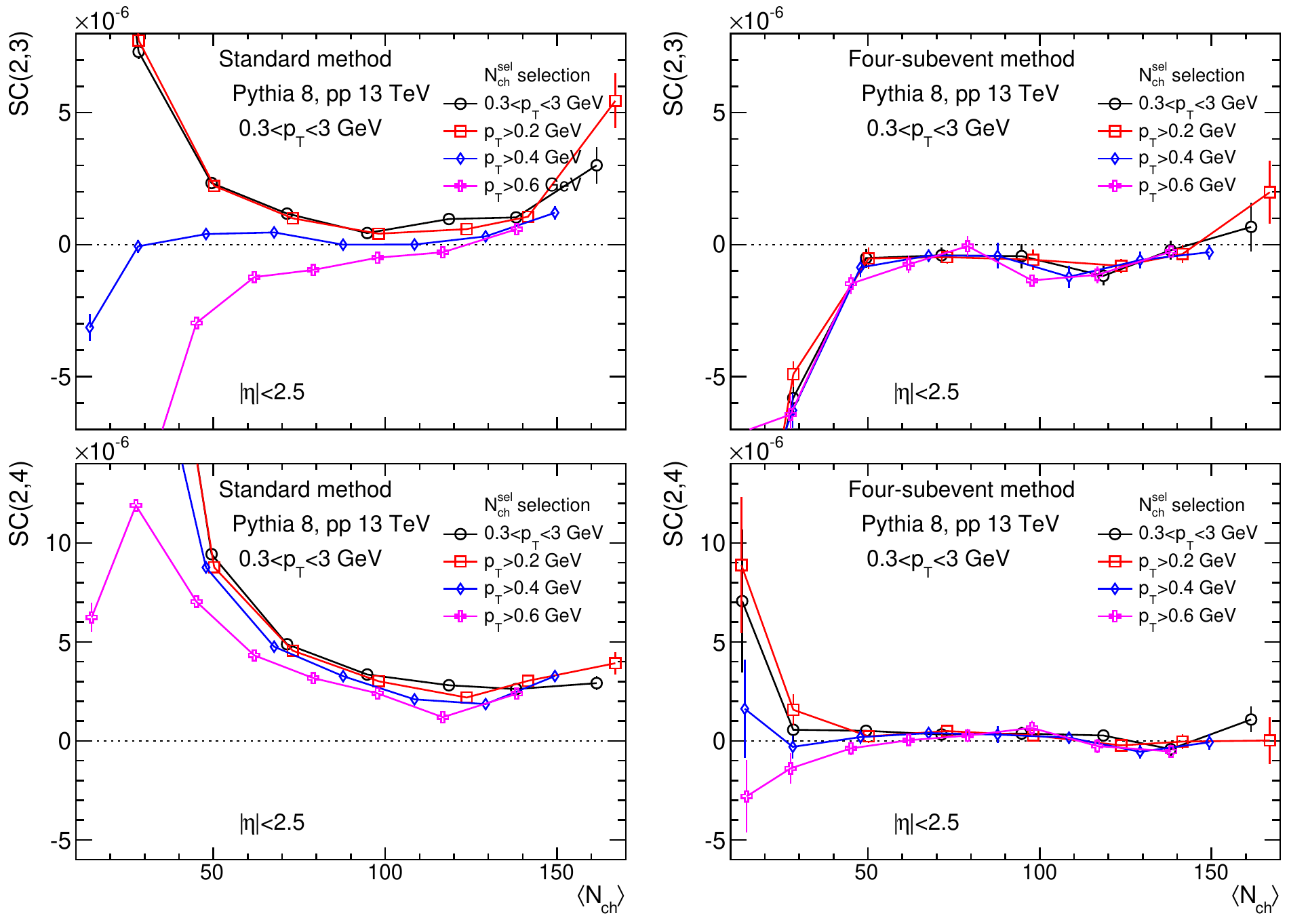}
\end{center}
\caption{\label{fig:1a} The $\scb$ (top row) and $\sca$ (bottom row) calculated for charged particles in $0.3<\pT<3$ GeV and several $\nchb$. They are obtained using the standard cumulant method (left column) and four-subevent method (right column) in $pp$ collisions generated with PYTHIA 8 model.}
\end{figure}

\begin{figure}[h!]
\begin{center}
\includegraphics[width=1\linewidth]{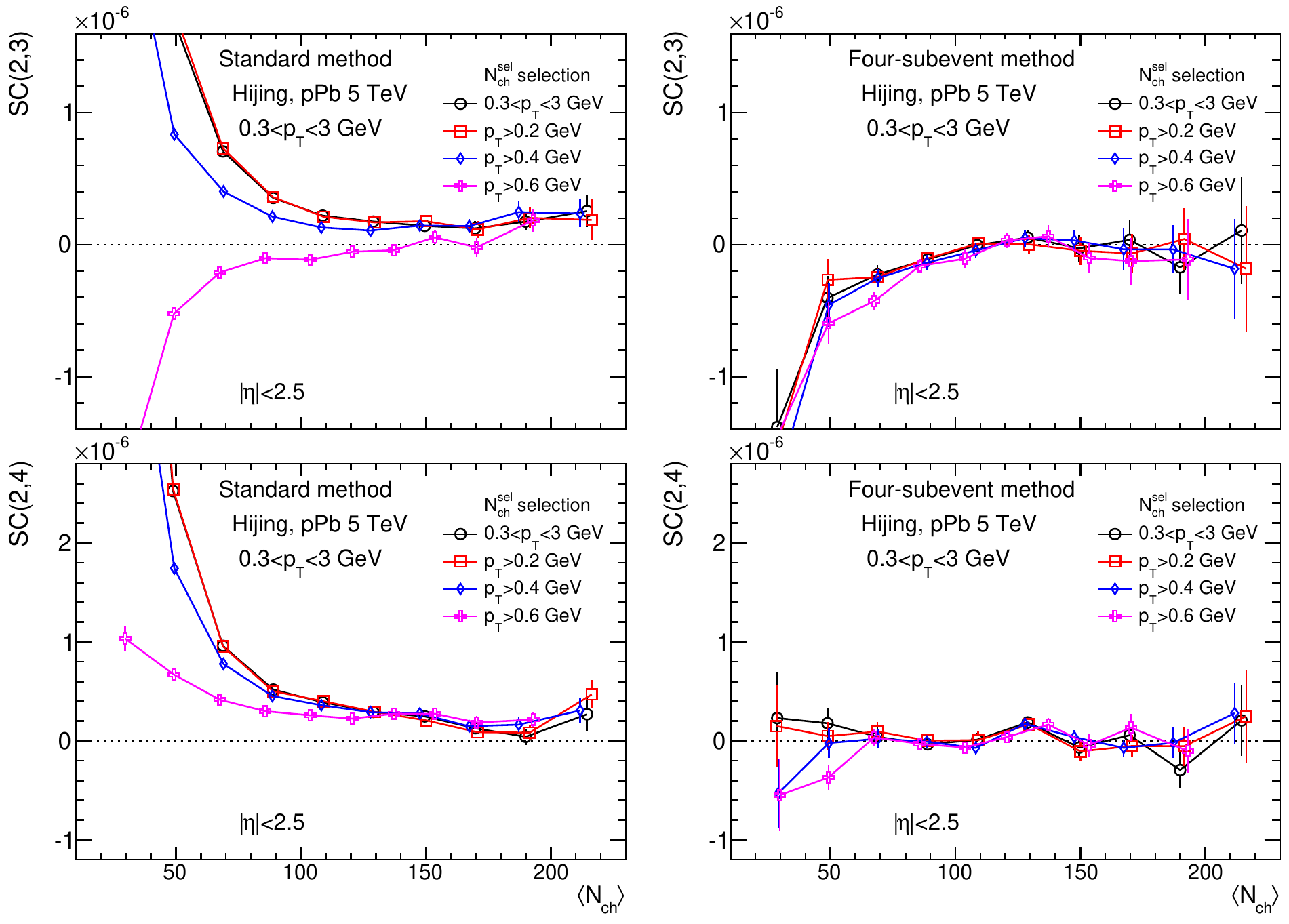}
\end{center}
\caption{\label{fig:2a}  The $\scb$ (top row) and $\sca$ (bottom row) calculated for charged particles in $0.3<\pT<3$ GeV and several $\nchb$. They are obtained using the standard cumulant method (left column) and four-subevent method (right column) in $\pPb$ collisions generated with HIJING model.}
\end{figure}

\bibliography{cumu3_v2}{}
\bibliographystyle{apsrev4-1}

\end{document}